\documentclass[12pt]{article}
\usepackage{amsfonts,amsmath}
\usepackage{amssymb}

\usepackage[hypertex] {hyperref}
\makeatletter \@addtoreset{equation}{section} \makeatother

\newcommand{\be}{\begin{equation}}
\newcommand{\ee}{\end{equation}}
\newcommand{\bee}{\begin{eqnarray}}
\newcommand{\beee}{\begin{array}}
\newcommand{\eee}{\end{eqnarray}}
\newcommand{\eeee}{\end{array}}

 \newcommand{\bkap}{\overline{{\kappa}}}
\newcommand{\gt}{\tau}

\newcommand{\ga}{\alpha}
\newcommand{\pa}{{\dot{\ga}}}
\newcommand{\pb}{{\dot{\gb}}}
\newcommand{\pga}{{\dot{\gamma}}}
\newcommand{\gb}{\beta}
\newcommand{\gga}{\gamma}

\newcommand{\E}{{\cal E}}

\newcommand{\Hh}{{\cal H}}

\newcommand{\rhs}{{\it r.h.s.} }

\newcommand{\ie}{{\it i.e.,} }
\newcommand{\ls}{\!\!\!\!\!\!}

\newcommand{\gvep}{\varepsilon}

\newcommand{\go}{\omega}
\newcommand{\WW}{\mathcal{W} }
\newcommand{\WWW}{ {\WW}'{} }

\newcommand{\eq}{\eqref}

 \newcommand{\etc}{{\it etc}  }
\newcommand{\hhmt}{{{h}}}

{}
{}
{}

\newcommand{\SSS}{{\cal E}^0}
\newcommand{\BS}{{\cal E}^1}

\newcommand{\pp}{{\mathbf{p} }}

\newcommand{\drZ}{{{\rm d}_Z}}

\newcommand{\va}{{\bf a}}

\newcommand{\halfi}{\f{i}{2}}

\newcommand{\hmtop}{{\hat{\vartriangle}}}

\newcommand{\hmt}{{\vartriangle}}
\newcommand{\hmtt}{\hmt} 

\newcommand{\goo}{{w}}

\newcommand{\by}{{\bar{y}}}

 \newcommand{\bjc}{{ {j_{c}}}}

\newcommand{\q}{\,,\qquad}

\newcommand{\nn}{\nonumber}

\newcommand{\half}{\frac{1}{2}}

\newcommand{\p}{\partial}
\newcommand{\D}{{\cal D}}

\newcommand{\f}{\frac}

\newcommand{\shi}{{\rm s}}

\newcommand{\bu}{\bar{\kappa}}

\newcommand{\PP}{ \mathcal{P} }
\newcommand{\PPP}{ {J} }

\newcommand{\dgb}{{\dot \gb}}
\newcommand{\dga}{{\dot \ga}}

\newcommand{\dr}{{\rm d}}

\begin{document}

\begin{flushright}

{\small FIAN/TD/09-2018}
\end{flushright}
\vspace{1.7 cm}

\begin{center}
{\large\bf Homotopy Operators and Locality Theorems in\\
 Higher-Spin Equations}

\vspace{1 cm}

{\bf O.A.~Gelfond$^{1,2}$ and  M.A.~Vasiliev$^1$}\\
\vspace{0.5 cm}
\textbf{}\textbf{}\\
 \vspace{0.5cm}
 \textit{${}^1$ I.E. Tamm Department of Theoretical Physics,
Lebedev Physical Institute,}\\
 \textit{ Leninsky prospect 53, 119991, Moscow, Russia}\\

\vspace{0.7 cm} \textit{
${}^2$ Federal State Institution "Scientific Research Institute of System Analysis
of  Russian Academy of Science",\\
Nakhimovsky prospect 36-1, 117218, Moscow, Russia
}

\end{center}

\vspace{0.4 cm}

\begin{abstract}
\noindent
     A new class of shifted homotopy operators in higher-spin gauge theory
      is introduced. A  sufficient
     condition for locality of dynamical equations is formulated and Pfaffian Locality
     Theorem identifying a
 subclass of shifted homotopies that decrease the degree of  non-locality in higher orders
of the perturbative expansion is proven.

\end{abstract}


\newpage

\section{Introduction}
Nonlinear field equations for $4d$ massless fields of all spins  were
found in \cite{Vasiliev:1990en,Vasiliev:1992av}. The most symmetric vacuum solution to
 these equations describes $AdS_4$. Due to the presence  of $AdS_4$  radius as a
  dimensionful parameter, higher-spin (HS) interactions can contain
infinite tails of higher-derivative terms. This can make the theory non-local in the standard
sense, raising the question which field variables lead to the local or minimally non-local
setup in the perturbative analysis. Recently, in \cite{Vasiliev:2016xui,Gelfond:2017wrh,Vasiliev:2017cae}
it was shown how nonlinear HS equations of \cite{Vasiliev:1992av} reproduce current interactions
in the lowest order in interactions. It has been then checked in \cite{Sezgin:2017jgm,Didenko:2017lsn,Misuna:2017bjb}
that the results of \cite{Vasiliev:2016xui,Gelfond:2017wrh} properly reproduce the holographic
expectations thus resolving some of the puzzles of the
analysis of HS holography conjectures of \cite{Klebanov:2002ja,Aharony:2011jz,Giombi:2011kc}
encountered in  \cite{Giombi:2009wh,Giombi:2011kc,Giombi:2012ms}
(and references therein).

The derivation of \cite{Vasiliev:2016xui,Gelfond:2017wrh} was based on the separation of variables
(holomorphic factorization)
in the zero-form sector of the $4d$ HS theory.  So far the perturbative analysis of HS equations was
based on the {\it conventional} homotopy operator technics proposed in \cite{Vasiliev:1992av}.
In \cite{Vasiliev:2017cae} it was explicitly checked that, in agreement with
\cite{Giombi:2009wh,Boulanger:2015ova}, the field redefinition that brings the results
obtained by virtue of the  conventional homotopy to the correct local form is non-local. Moreover,
in  \cite{Vasiliev:2017cae} it was shown that from the perspective of the full nonlinear HS equations
the field redefinition found in \cite{Vasiliev:2016xui} has distinguished properties indicating that
it  leads to minimal order of non-locality in the higher orders.
However  it was not clear how  the homotopy technics should be  modified to lead directly to
the correct local results in the perturbative analysis of HS equations with no reference to field
redefinitions.

The main aim of this paper
is to generalize the  conventional homotopy technics in such a way that it will give immediately correct local
results in the lowest order.
Based on this generalization we  prove a  theorem showing how to choose the proper class
of homotopy operators to decrease the level of non-locality of HS equations in  higher orders as well.

Note that what is interpreted as locality  in this paper
is probably better to call spin locality as it refers to the form of expressions in
the sector of spinor variables underlying the unfolded formulation of HS equations of
\cite{Vasiliev:1990en,Vasiliev:1992av}. Its relation to the conventional definition in
terms of space-time derivatives is via unfolded equations as we briefly
recall now.

Unfolded   equations   of
$4d$  massless   Fronsdal \cite{Frhs,Frfhs}  fields of all spins
$s=0,1/2,1,3/2,2\ldots $  in $AdS_4$   are formulated in terms of  a
{one-form} $  \omega (Y;K| x)= dx^n\omega_n (Y;K| x)$ and zero-form $ C(Y;K| x)  $ \cite{Ann}, $Y=(y,\by)$.
The  Klein operators $K=(k,\bar k)$ satisfy
\be
\label{kk}
k y^\ga = -y^\ga k\,,\quad
k \bar y^\pa = \bar y^\pa k\,,\quad
\bar k y^\ga = y^\ga \bar k\,,\quad
\bar k \bar y^\pa = -\bar y^\pa \bar k\q kk=\bar k\bar k = 1\,,\quad
k\bar k = \bar k k\,.
\ee
More precisely, to describe massless fields, the one-form $  \omega (Y;K| x)$
should be even in $k,\bar k$  while the zero-form $C (Y;K| x)$
should be odd.
Thus, massless fields are doubled
\be\label{Csumkbark}
C(Y;K|x)= C^{1,0}(Y|x) k + C^{0,1}(Y|x) \bar k\q \go(Y;K|x)= \go^{0,0}(Y|x)  + \go^{1,1}(Y|x) k\bar k\,.
\ee

Unfolded  field equations
for free massless
fields of all spins in the $AdS_4$ are  \cite{Ann}
\bee
\label{CON1}
    && \ls\ls\ls R_1(Y;K| x) = L(w,C):=
\f{i}{4} \!\Big (\! \eta \overline{H}^{\dga\pb}\f{\p^2}{\p \overline{y}^{\dga} \p \overline{y}^{\dgb}}\
{ C}(0,\overline{y};K| x) k +\bar \eta H^{\ga\gb}\! \f{\p^2}{\p
{y}^{\ga} \p {y}^{\gb}}\!
{C}(y,0;K| x) \bar k\!\Big ) , \quad\\\label{CON2}
\,&& \qquad\tilde{D}C (Y;K| x) =0\,, \eee
where
\be
\label{RRR}
R_1 (Y;K| x) :=D^{ad}\omega (Y;K |x) :=
D^L \go  (Y;K| x) +
\lambda h^{\ga\pb}\Big (y_\ga \frac{\partial}{\partial \bar{y}^\pb}
+ \frac{\partial}{\partial {y}^\ga}\bar{y}_\pb\Big )
\omega  (Y;K | x) \,,
\ee
\be
\label{tw}
\tilde D C(Y;K |x) :=
D^L C (Y;K |x) -{i}\lambda h^{\ga\pb}
\Big (y_\ga \bar{y}_\pb -\frac{\partial^2}{\partial y^\ga
\partial \bar{y}^\pb}\Big ) C (Y;K |x)\,,
\ee
\be
\label{dlor}
D^L f (Y;K|x) :=
\dr_x f (Y;K|x) +
\Big (\go_L^{\ga\gb}y_\ga \frac{\partial}{\partial {y}^\gb} +
\overline{\go}_L^{\pa\pb}\bar{y}_\pa \frac{\partial}{\partial \bar{y}^\pb} \Big )f (Y;K|x)\q
\dr_x: =dx^n\f{\p}{\p x^n}
\,.
\ee
Background $AdS_4$ space of radius $\lambda^{-1}=\rho$
is described by a flat $sp(4)$
connection $w=(w_{\alpha \gb},\overline{w}_{\dga\dgb},h_{\ga\dgb})$
containing Lorentz connection
$w_{\alpha \gb} $, $\overline{w}_{\dga\dgb}$ and
vierbein  $h_{\ga\dgb}$ that obey
\be
\label{nR}
\dr_x w_{\alpha \gb} +w_{\alpha}{}_\gamma
 w_{\gb}{}^{\gamma} -\lambda^2\, H_{\alpha \gb}=0\,,\quad
\dr_x\overline{w}_{{\pa}
{\pb}} +\overline{w}_{{\pa}}{}_{\dot{\gamma}}
 \overline{w}_{{\pb}}{}^{ \dot{\gga}} -\lambda^2\,
 \overline H_{{\pa\pb}}=0\,,\quad
\dr_x h_{\alpha{\pb}} +w_{\alpha}{}_\gamma
h^{\gamma}{}_{\pb} +\overline{w}_{{\pb}}{}_{\dot{\delta}}
 h_{\alpha}{}^{\dot{\delta}}=0\,,
\end{equation}
where $H^{\ga\gb} := h^{\ga}{}^\pa  h^\gb{}_\pa$ and $\overline{H}^{\pa\pb} :=
h^{\ga}{}^\pa h_{\ga}{}^{\pb}$ are the frame two-forms (wedge symbol is  omitted).

In the massless sector, system (\ref{CON1}), (\ref{CON2}) decomposes
into  subsystems of different spins, with a  spin $s$  described by
the one-forms $ \omega (y,\bar{y};K| x)$ and zero-forms $C (y,\bar{y};K| x)$ obeying
\be
\label{mu}
\omega (\mu y,\mu \bar{y};K\mid x) = \mu^{2(s-1)} \omega (y,\bar{y};K\mid x)\q
C (\mu y,\mu^{-1}\bar{y};K\mid x) = \mu^{\pm 2 s}C (y,\bar{y};K\mid x)\,,
\ee
where  $+$ and $-$   correspond to helicity $h=\pm s$ selfdual and anti-selfdual parts
of the generalized Weyl tensors $C (y,\bar{y};K| x)$.
For spins $s\geq 1$, equation (\ref{CON1})
expresses the Weyl {0-forms} $C(Y;K|x)$ via
gauge invariant combinations of derivatives of the HS gauge connections.
More precisely, the primary-like Weyl {0-forms} are just the holomorphic and antiholomorphic
parts $C(y,0;K|x)$ and $C(0,\bar y;K|x)$ which appear on the
\rhs of Eq.~(\ref{CON1}).
Those associated with higher powers of auxiliary variables
$y$ and $\bar y$ describe on-shell nontrivial combinations of derivatives
of the generalized Weyl tensors as is obvious from Eqs.~(\ref{CON2}), (\ref{tw})
relating second derivatives in $y,\bar y$ to
the $x$ derivatives of  $C (Y;K|x)$ of lower degrees
in $Y$. Hence higher derivatives in the nonlinear system hide in the
 components of $C (Y;K| x)$ of higher orders in $Y$. To see whether the
 resulting equations are local or not at higher orders one has to inspect
 the dependence of vertices on the higher components of   $C (Y;K| x)$.

At the linearized level, Eq.~(\ref{tw}) implies that $\f{\p}{\p x}$ is equivalent
to $\f{\p^2}{\p y\p\bar y}$. Hence,  at this level the analysis of spin locality in terms of
$y,\bar y$ variables is equivalent to that in terms of space-time derivatives. However in
higher  orders  Eq.~(\ref{tw}) acquires  nonlinear corrections. This makes
the relation between spin locality in terms of $y,\bar y$ variables and space-time locality
less straightforward. Since the spinor sector of HS equations is of fundamental
importance   all concepts in HS theory including locality have to be originally
defined in these terms. Therefore, we regard the spin locality of the HS theory as the
fundamental concept. Relation to the space-time locality at higher orders is not
straightforward being somewhat analogous to the effect of current exchange contribution
in the space-time formulation.

 A related comment is that space-time covariant derivatives
$D^L$ do not commute in presence of non-zero cosmological constant which is of order one in
the absence of other dimensionful parameters like $\alpha'$ in string theory.
This raises a nontrivial question of the  choice of the ordering prescription
in the expansions in higher space-time derivatives. We believe that the concept of spin
locality in terms of spinor variables provides an appropriate solution to this problem
which may be hard to guess directly in the space-time approach.

Let us explain the idea of the analysis of spin locality in some more detail.
As explained in Section \ref{Locality lemma},
general exponential representation for the order-$n$ corrections in
the zero-forms $C$ is
\be
\label{C.C}
\sum_{\pp\bar \pp}\int d\tau \hat {\mathcal P}^{\pp\bar \pp}_{n}(y,\bar y,p,\bar p,\tau)
\hat E^{\pp\bar \pp}_{n}(y,\bar y,p,\bar p,\tau)C(Y_1;K)\ldots C(Y_n;K)\big|_{Y_j=0}\,,
\ee
where
\be
\label{TailorC}
 p^j{}_\ga := - i\f{\p}{\p y_j^\ga}\q  \bar p^j{}_\pa :=  -i\f{\p}{\p \by_j^\pa} \,,
 \ee
 $ \hat {\mathcal P}^{\pp\bar \pp}_{n}(y,\bar y,p,\bar p,\tau)$ is some polynomial of $y,\bar y$
$p^i$ and $\bar p^i$ with coefficients being
regular functions of some homotopy integration parameters $\tau$ and
\be
\hat E^{\pp\bar \pp}_{n }= \hat E^{\pp}_{n}\hat {\bar E}^{\bar \pp}_{ n}\q
\label{Enh} \hat E^\pp_n(\hat B,\hat P,p|z,y)=
\exp  i (- \hat B_j(\tau) p^j_\ga y^\ga
+ \half \hat P_{ij} (\tau) p^i{}^\ga p^j{}_\ga ) \,
k^\pp\,,
\ee
where $\pp=0,1$ and parameters
 $\hat B  \in   \mathbb{C}^n$, $\hat P_{ij}=-{\hat P}_{ji}\in\mathbb{C}^n\times\mathbb{C}^n$
  may be $\tau$-dependent.

Spin locality of HS interactions is governed by the
coefficients $\hat P_{ij}$ in $\hat E^\pp_n$ (\ref{Enh}) and their complex conjugates
$\hat {\bar P}_{ij}$ in $\hat {\bar E}^{\bar \pp}_n$ that determine contractions between,
respectively, undotted and dotted spinorial arguments of different factors of $C(Y;K|x)$.
Since the contribution of $\hat P_{ij}$ and $\hat {\bar P}_{ij}$-dependent terms is via
exponential it gives rise to a non-polynomial expansion in $p^i{}^\ga p^j{}_\ga$ and
$\bar p^i{}^\dga \bar p^j{}_\dga$ and,  hence, via  (\ref{CON2}) and (\ref{tw}), to non-local
expansion in space-time derivatives. In fact, nonlinear corrections to HS equations have
the form (\ref{C.C}), (\ref{Enh}) where at least one of the coefficients  $\hat P_{ij}(\tau)$
and $\hat {\bar P}_{ij}(\bar \tau)$ is nonzero. This is a manifestation of the fact
 that HS theory is in a certain sense non-local in agreement with
 the well-known property that higher spins carry higher derivatives and, hence, in presence of an
 infinite tower of HS fields the full theory must contain infinite tower of higher derivatives
 as well.

 A less trivial question is on the locality of vertices involving particular
 spins $s_1,\ldots s_n$. In accordance with (\ref{mu}), for fixed helicities,
 the degree in $y_i$ variables in $C(Y_i;K|x)$ is related to that in  $\bar y_i$.
 In that case the degree in $ p^i{}^\ga p^j{}_\ga$ gets related to that in
 ${\bar p}^i{}^\dga {\bar p}^j{}_\dga$ in a particular vertex. As a result, for vertices with fixed spins
 polynomiality in $ p^i{}^\ga p^j{}_\ga$
 implies polynomiality in ${\bar p}^i{}^\dga {\bar p}^j{}_\dga$ and vice versa.
 Hence spin locality for a any fixed set of spins will be achieved if, for instance,
 one of the coefficients $\hat P_{ij}$ or $\hat {\bar P}_{ij}$ is zero. If it happened in
 all orders, this would imply all-order spin locality of HS equations.

One of the main results of this paper consists of the proof of {Pfaffian Locality Theorem}
in Section \ref{Pfaffian Locality Theorem} stating that, in the sector of  equations on zero-forms,
 there exist such particular homotopy procedures  that the antisymmetric  matrices $\hat P_{ij}(\tau)$ and $\hat {\bar P}_{ij}(\bar \tau)$ are degenerate. For the lowest order
 bilinear corrections associated with $2\times2 $ antisymmetric matrices this implies that
 at least one of the   matrices $\hat P_{ij}$ and $\hat {\bar P}_{ij}$ is zero thus implying
 spin locality of the lowest-order corrections.
In fact, this result allows us to speculate that, by  a proper choice of homotopy operators, HS contact interactions can be brought to the local form in higher-orders for every fixed set
of spins in a vertex. (Note that some higher-order local vertices are constructed in \cite{DGKV}.)
If nevertheless this does not happen then it makes sense to look for a minimally non-local setup
such that,
being non-local, it is minimally non-local  leading to the fastest decrease of the coefficients in front of higher
powers of $ p^i{}^\ga p^j{}_\ga$ and ${\bar p}^i{}^\dga {\bar p}^j{}_\dga$.
Pfaffian Locality Theorem  indicates that such minimization should be possible.

Let us stress that there are many reasons why it is important
to elaborate the intrinsic analysis of the
HS gauge theory in the bulk with no reference to the holographic duals.
The simplest is that apart from free boundary theories dual to particular
HS gauge theories in the bulk, the latter equally well describe interacting
Chern-Simons boundary theories \cite{Aharony:2011jz,Giombi:2011kc}  where the computation of amplitudes is  more
involved.  More general background solutions of HS theories with more complicated boundary
 duals  like for instance massive deformations  can also be of  interest. Another point is that the approach
 proposed in this paper is applicable to a much more general class
 of Coxeter HS theories some of which were conjectured in \cite{Vasiliev:2018zer}
 to be related to String Theory upon spontaneous breakdown of HS symmetries.
 In the latter case, application of holographic duality is more tricky because
 it becomes strong-weak duality. Hence, independent formulation of the underlying
 bulk HS theory is of great importance in that case as well.

 The rest of the paper is organized as follows.
 The form of nonlinear HS equations is
 sketched in Section \ref{Nonlinear Higher-Spin Equations}. Perturbative analysis
 of HS equations in terms of homotopy operator technics is recalled in Section \ref{peran}.
 In Section \ref{Shift homotopy}
 we  introduce modified homotopy operators appropriate for the analysis of locality
 of HS equations.  In Section \ref{Locality lemma} we prove $Z$-dominance Lemma
 providing a sufficient criterion for the locality of nonlinear corrections to dynamical field
 equations.   Pfaffian Locality Theorem providing  a criterion for the choice of homotopy
 decreasing  the degree of non-locality in higher orders of interactions is proven in Section
 \ref{Pfaffian Locality Theorem}.  Section \ref{conc} contains brief conclusions.

\section{Nonlinear higher-spin equations}
\label{Nonlinear Higher-Spin Equations}
$4d$ nonlinear HS equations \cite{Vasiliev:1992av} have the form
\bee\label{eq:HS_1}
 &  & \mathrm{d}_{x}\WW+\WW*\WW= i (\theta^A \theta_A + F_*(B)* \gamma +
 \bar F_*(B)* \bar\gamma  ) \,, \\
 &  & \mathrm{d}_{x}B+\WW*B-B*\WW=0\,\label{eq:HS_2}\,,
\eee
where
\be\label{gamma=} \gamma=\theta^\ga \theta_\ga  \kappa k\q
\bar\gamma=\bar \theta^\pa \bar \theta_\pa \bkap \bar k\,
.\ee
$\WW$ and $B$ are fields of the theory which depend both on space-time
coordinates $x^n$ and on twistor-like variables $Y^{A}=\left(y^{\alpha},\bar{y}^{\dot{\alpha}}\right)$
and $Z^{A}=\left(z^{\alpha},\bar{z}^{\dot{\alpha}}\right)$.  ($A=1,\ldots 4$ is a Majorana spinorial index
while $\ga = 1,2$ and $\dga =1,2$ are two-component ones. The latter are raised and lowered
by $\varepsilon_{\ga\gb}=-\varepsilon_{\gb\ga}$, $\varepsilon_{12}=1$: $A^\ga =\gvep^{\ga\gb} A_\gb$,
$A_\ga = A^\gb\gvep_{\gb\ga}$ and analogously for dotted indices.)

The $Y$ and $Z$ variables provide a realization of HS algebra through
the following noncommutative  associative star product $*$ acting on functions of two
spinor variables
\be
\label{star2}
(f*g)(Z;Y)=
\int \f{d^{4} U\,d^{4}V}{(2\pi)^{4}}  \exp{[iU^A V^B C_{AB}]}\, f(Z+U;Y+U)
g(Z-V;Y+V) \,,
\ee
where
$C_{AB}=(\epsilon_{\ga\gb}, \bar \epsilon_{\dga\dgb})$
is the $4d$ charge conjugation matrix and
$ U^A $, $ V^B $ are real integration variables.
 1 is a unit element of the star-product
algebra, \ie $f*1 = 1*f =f\,.$ Star product
(\ref{star2}) provides a particular
realization of the Weyl algebra
\be
[Y_A,Y_B]_*=-[Z_A,Z_B ]_*=2iC_{AB}\,,\qquad
[Y_A,Z_B]_*=0\q [a,b]_*:=a*b-b*a\,.
\ee

    It is convenient to introduce anticommuting $Z-$differentials $\theta^A$, $\theta^A \theta^B=-\theta^B
\theta^A$.  \textbf{$B$ }is a 0-form, while $\WW$ is differential one-form
 with respect to $dx^n$, $\theta^A$ differentials, \ie $\WW=\{W\,,S\}$, where
$W(Z;Y;K|x)$ is a space-time one-form, while
$
S=\theta^A S_A (Z;Y;K|x) \,.
$

The  Klein operators satisfy relations analogous to (\ref{kk})
 with
 $y^\ga\to w^\ga= (y^\ga, z^\ga, \theta^\ga )$, $\bar y^\dga\to\bar w^\pa =
(\bar y^\pa, \bar z^\pa, \bar \theta^\pa )$, which extend the action of the star product to the
Klein operators.
Decomposing master-fields  with respect to the Klein-operator parity,
 $A^\pm(Z;Y;K|x)=\pm A^\pm(Z;Y;-K|x)$, HS gauge fields are
 $W^+,S^+$ and $B^-$ while  $W^-$, $S^-$ and $B^+$
 describe an infinite tower of topological fields
 with every $AdS_4$ irreducible field describing at
most a finite number of degrees of freedom. (For more detail see
\cite{Vasiliev:1992av,Vasiliev:1999ba}).

 $F_*(B) $ is some star-product function of the field $B$.
The simplest choice  of the linear function
$
F_*(B)=\eta B$, $\bar F_* (B) = \bar\eta B\,,
$
where $\eta$ is a complex parameter
$
\eta = |\eta |\exp{i\varphi}$, $\varphi \in [0,\pi)\,,
$
leads to a class of pairwise nonequivalent nonlinear HS
theories. The  cases
of $\varphi=0$ and $\varphi =\f{\pi}{2}$  correspond
to so called $A$ and $B$ HS models
 that   respect parity \cite{Sezgin:2003pt}.

The left and right inner Klein operators
\be
\label{kk4}
\kappa :=\exp i z_\ga y^\ga\,,\qquad
\bu :=\exp i \bar{z}_\dga \bar{y}^\dga\,,
\ee
 which enter Eq.~\eq{gamma=}, change a sign of
 undotted and dotted spinors, respectively,
\be
\label{uf}
\!(\kappa *f)(z,\!\bar{z};y,\!\bar{y})\!=\!\exp{i z_\ga y^\ga }\,\!
f(y,\!\bar{z};z,\!\bar{y}) ,\quad\! (\bu
*f)(z,\!\bar{z};y,\!\bar{y})\!=\!\exp{i \bar{z}_\dga \bar{y}^\dga
}\,\! f(z,\!\bar{y};y,\!\bar{z}) ,
\ee
\be
\label{[uf]}
\kappa *f(z,\bar{z};y,\bar{y})=f(-z,\bar{z};-y,\bar{y})*\kappa\,,\quad
\bu *f(z,\bar{z};y,\bar{y})=f(z,-\bar{z};y,-\bar{y})*\bu\,,
\ee
\be
\kappa *\kappa =\bu *\bu =1\q \kappa *\bu = \bu*\kappa\,.
\ee

\section{Perturbative analysis and homotopy operator}
\label{peran}
\subsection{Vacuum}

Perturbative analysis of  Eqs.~(\ref{eq:HS_1}), (\ref{eq:HS_2}) assumes their
linearization around some vacuum solution.
The simplest choice is
\be
W_0(Z;Y;K|x)= w(Y;K|x)\q S_0(Z;Y;K|x) = \theta^A Z_A\q B_0(Z;Y;K|x)=0\,,
\ee
where $w(Y|x)$ is some solution to the flatness condition
\be
\label{flat}
\dr_x w + w*w=0\,.
\ee
A flat connection $w(Y|x)$, that describes $AdS_4$ via (\ref{nR}), is bilinear in $Y^A$
\be
\label{ads}
w(Y|x) =  -\f{i}{4} (\goo^{\ga\gb}(x) y_\ga y_\gb + \bar \goo^{\dga\dgb}(x)\bar y_\dga \bar y_\dgb
+2h^{\ga\dgb}(x) y_\ga \bar y_\dgb )\,.
\ee

Since $S_0$ has a trivial star-commutator with the Klein operators, a simple computation
gives
\be
[S_0\,, F(Y;Z;K|x)]_* = -2i \dr_Z  F(Y;Z;K|x)\q
\dr_Z = \theta^A \f{\p}{\p Z^A}\,.
\ee

Let \be\WW(Y,Z|x)=S_0+\goo(Y|x) +\WWW(Y,Z|x)\,.
\ee
Denoting
\be\label{Dkrugloe}
\D:=  \D_x+\D_\goo  \q \D_x=- \halfi   \dr_x\q     \D_\goo A:=-\halfi [\goo\,,A]_{\pm*}
\,,\ee
 Eqs.~(\ref{eq:HS_1}), (\ref{eq:HS_2}) yield
 \bee\label{GENequatons}
\left(\drZ -\D \right) \WWW+\halfi \WWW*\WWW
&=& -\half  (  \eta   B*  \gamma +  \bar \eta
  B * \bar\gamma)\q\\ \label{2}
\left(\drZ -\D \right) B&=&-\halfi [\WWW\,,B]_* \,.\qquad
\eee

\subsection{Homotopy trick}
\label{Homotopy trick}
To eliminate $Z$-variables 
one has to  repeatedly solve equations of the form
\be
\label{fg}
\dr_Z f(Z;Y;K|x) = J(Z;Y;K|x)\,.
\ee
  Consistency of  HS equations    guarantees that
  $J(Z;Y;K|x)$ is $\dr_Z$-closed,
 $\dr_Z J(Z;Y;K|x)=0\,,$
 implying formal consistency of Eq.~(\ref{fg}). However, it admits a solution
 only if $J$ is $\dr_Z$-exact.

 Given homotopy operator $\partial$
\begin{equation}
\partial^{2}=0\,,\label{eq:dwdw_0}
\end{equation}
the operator
\begin{equation}
A:=\{\mathrm{d}_Z\,,\partial\}\label{eq:A_d_d}
\end{equation}
obeys
\begin{equation}
[\mathrm{d}_Z\,,A]=0\,,\qquad[\partial\,,A]=0\,.
\end{equation}
For diagonalizable
$A$, the standard Homotopy   Lemma states that cohomology $H_{\mathrm{d}_Z}$ of $\mathrm{d}_Z$,
 is  in the kernel of $A$
\begin{equation}
H_{\mathrm{d}_Z} \subset KerA\,.\label{eq:hom_lemma}
\end{equation}
In this case the projector ${h}$ to $KerA$
\begin{equation}
{h}^{2}={h}
\end{equation}
and  the operator $A^{*}$ can be defined to obey
\begin{equation}
[{h}\,,\mathrm{d}_Z]=[{h}\,,\partial]=0\q
A^{*}A=AA^{*}=Id-{h}\,.\label{eq:AA*_Id}
\end{equation}
The {\it resolution operator}
\be\label{standres}
\hmt:=A^{*}\partial=\partial A^{*}
\ee
gives
 the resolution of identity
\be
\left\{ \mathrm{d}_Z\,,\hmt\right\} +{h}=Id\,\label{eq:res_id_gen}
\ee
allowing to find a  solution to the equation
$\mathrm{d}_Zf=J$
with $\mathrm{d}_Z$-closed $J$ outside of $H_{\mathrm{d}_Z}$,
\ie obeying  $\hat{h}J=0$, in the form
\begin{equation}
f=\hmt J+\mathrm{d}_Z\epsilon+g,\label{eq:gen_sol}
\end{equation}
where an exact part $\mathrm{d}_Z\epsilon$  and $g\in H_{\mathrm{d}_Z}$
remain undetermined.

\subsection{Perturbative expansion}
\label{Perturbations}

    HS equations reconstruct the  dependence on
 $Z_A$ in terms of the zero-form
$ C(Y;K|x)\in H_{\dr_Z}$ and one-form $ \go(Y;K|x)\in H_{\dr_Z} $ representing the
$\dr_Z$-cohomological parts  of $B$ and $\WWW$,
\be
\label{ijda}B=C(Y;K|x)+ \sum_{j=2}^\infty B_j(Y,Z;K|x)\q
\WWW =\go(Y;K|x)+ \sum_{j=1}^\infty \WW_j(Y,Z;K|x) \,,\qquad
\ee
where   zero-forms
$ B_j(Y,Z;K|x)$ and  one-forms $ \WW_j(Y,Z;K|x)$
are of order $j$  in  $\go$ and $C$   and obey
 \be
  H_{\dr_Z}\big(B_j(Y,Z;K|x)\big)=0\q H_{\dr_Z}\big(\WW_j(Y,Z;K|x)\big)=0\,\qquad \forall j.\ee
The perturbative analysis goes as follows.
 Suppose that an order-$n$ solution
   \bee\label{WW=}  \WW'{}^{(n)}(Y,Z;K|x)&=&      \go(Y ;K|x)+\sum_{j=1}^n \WW_j(Y,Z;K|x)
    \q\\ \label{B=}
  B^{(n)}(Y,Z;K|x)&=&   \sum_{j=1}^n B_j(Y,Z;K|x)\,\q  B_1(Y,Z;K|x)=C(Y ;K|x)
  \eee
is  found. Eqs.~\eq{GENequatons}, \eq{2}     yield at   order $ n+1$
  \bee\label{GENequaton+1sn+1}
\Big(\left(\drZ -\D \right) \WW'{}^{(n+1)}\Big)\Big|_{\le n+1}
\!\!\!\!&\!\!=\!\!&\!\!-\half\Big( i \WW'{}^{(n+1)}*\WW'{}^{(n+1)}
\!+\!  \eta  B^{(n+1)}*\!  \gamma \!+\!   \bar \eta
  B^{(n+1)} * \bar\gamma\!\Big)\Big|_{\le n+1}+\cdots \q\quad\\ \nn
\Big(\left(\drZ -\D \right) B^{(n+1)}\Big)\Big|_{\le n+1}
\!\!\!\!&\!\!=\!\!&\!\!-\half\Big( i  [\WW'{}^{(n+1)}\,,B^{(n+1)}]_*
\Big)\Big|_{\le n+1} +\cdots\, ,\qquad
\eee
where ellipsis  denotes higher-order terms, $\left.   A(C,\go)\right|_{k}$
is the order-$k$ part  of $A(C,\go)$  in  $\go$ and $C$,
    and
\be\nn\left.  A(C,\go)\right|_{ \le m}:=\bigcup_{k\le m}\left.A(C,\go)\right|_{k}.
\ee
From \eq{GENequaton+1sn+1} it follows by virtue of \eq{WW=}, \eq{B=}
using  \eq{Dkrugloe}
  \bee\label{GENeqn+1W}
-\D\go+ \theta(n\!-\!2)\halfi \go*\go+ \sum_{m=1}^{n+1}     \drZ   \WW_m
\!\!&\!=\!&\!\D_x     \WW_{n+1}+\!\sum_{m=1}^{n+1}  \left\{ X^\WW_{m} +\D_\goo    \WW_m
+  \left(\! \D_x \sum_{j=1}^{m-1 }   \!\WW_j
   \right)\!\Big|_{ m} \! \right\}
   \!\!\!\q\quad\\ \label{GENeqn+1B}
-\D C+    \sum_{m=2}^{n+1}   \drZ    B_m
\!\!&\!=\!&\!\D_x     B_{n+1}+\!\sum_{m=2}^{n+1} \left\{  X^B_{m}+ \D_\goo    B_m
+\left(\!\D_x   \sum_{j=2}^{m-1}  B_j  \right)\!\Big|_{ m}  \right\}
\! ,\qquad
\eee
where
\bee\label{rhsmW}{X}^\WW_{m}\!
\! &\!=\!&\!- \halfi  \sum_{j=1}^{m-1 } \WW_j *\WW_{m-j}-\halfi \left\{\go \,,  \WW_{m-1} \right\}_*
-\half    \eta    B_{m} *\gamma \!-\!\half\bar \eta
   B_{m} * \bar \gamma
\q \qquad\\  \label{rhsmB}
{X}^B_{m}
&\!
=\!&\!
-\half   \sum_{j=1}^{m-1} [B_j \,,\WW_{m-j}]_*
.
\eee
Let us stress that being of order $m$ in $C$ and $\go$,
${X}^\WW_{m}$ and ${X}^B_{m}$  contain $B_j $ and $\WW_{ j}$ with $j<m$.
Also it is used that the order-$n$  parts of dynamical equations     are of the form
\be\label{dynschnW}
\dr_x \go(Y;K|x)=\sum_{j=1}^n  J^0_j(Y;K|x)\q
\qquad\dr_x C(Y;K|x) =\sum_{j=1}^n J^1_j(Y;K|x)
\,,
\ee
where the two-forms $J^0_j\in H_{\dr_Z}$  and one-forms $J^1_j\in H_{\dr_Z}$
 are of order-$j$   in   $\go$  and $C$.

Since, acting trivially on $\go$ and $C$, $\dr_Z$ does not mix different perturbation orders,
equations \eq{GENeqn+1W} and \eq{GENeqn+1B}
 are $\dr_Z$-closed separately at any $m$.
 This allows one to use different   homotopy   operators for any $m$    in each of these equations.

\section{Shifted homotopy}
\label{Shift homotopy}


The {\it conventional}  homotopy operator
 \begin{equation}
\label{p}
\partial=Z^{A}\frac{\partial}{\partial\theta^{A}}\,
\end{equation}
 and resolution
 \begin{equation}
\hmt J\left(Z;Y;\theta\right)=Z^{A}\dfrac{\partial}{\partial\theta^{A}}\intop_{0}^{1}dt\dfrac{1}{t}J\left(tZ;Y;t\theta\right)
\label{eq:dz*}
\end{equation}
  were used in the
perturbative analysis of HS equations since \cite{Vasiliev:1992av}. Though being
 simple and looking  natural, they are known to lead to non-localities beyond the free field level \cite{Giombi:2009wh,Boulanger:2015ova,Vasiliev:2017cae}.

An obvious freedom in the definition
of  homotopy operator (\ref{p}) is to replace $Z^A$ by $Z^A+ a^A$ with some $Z$-independent $a^A$,
\begin{equation}
\label{pp}
\partial\to \partial_a =(Z^{A}+a^A)\frac{\partial}{\partial\theta^{A}}\q
\frac{\partial}{\partial Z^{A}} (a^B)=0\,.
\end{equation}
Resolution $\hmt_{a}$  and cohomology projector $\hhmt_{a }$ act
on a  $\phi(Z,Y,\theta)$ as follows
\be\label{homint} \hmt_{ a} \phi(Z,Y, \theta) =\int_0^1 \f{dt}{t} (Z+ a)^A\f{\p}{\p \theta^A}
 \phi( tZ-(1-t) a,t\theta)\q\hhmt_{a } \phi(Z,Y, \theta)= \phi(-a,Y,0)\,.
\ee
 $\hmt_{0}$ is the conventional resolution \eq{eq:dz*}.
The resolution of identity has  standard form
\be\label{newunitres}
\left\{ \drZ\,,\hmt_{a }\right\} +\hhmt_{a }=Id\, .\ee

For instance, one can set $a^A= c Y^A$ with some constant  $c$.\footnote{Retrospectively,
one can see that the form of HS equations presented
in \cite{Vasiliev:1990en} results from such  modification of the homotopy operator
with $c=\pm1$. However,  the formulation of \cite{Vasiliev:1990en}
demanded some non-local field redefinition even at the linear order. This
 problem  was later resolved in \cite{Vasiliev:1992av} via
application of the conventional homotopy upon introduction of  $Z$-variables and
the fields $S_A$.}
Naively,  this exhausts all Lorentz covariant choices for  $a^A$.
This  is however not  the case since  $a^A$ can also
be composed from derivatives with respect to arguments of $ \go(Y;K)$ and $C(Y;K)$
in $J=J(\go,C)$ in  (\ref{fg}).

Let
\be\label{dinfilord}  \Phi^1(Y;K) = \go (Y;K)\q\Phi^0(Y;K) =C(Y;K).\ee
Various terms on the \rhs of HS field equations  contain  ordered
products
\be
\label{ord}
\ls\Phi_n^{\va}(Y ;K) = \Phi^{a_1}(Y_1;K)\Phi^{a_2}(Y_2;K)\ldots\Phi^{a_n}(Y_n;K)\big|_{Y_i=Y},\quad
\va=\{a_1,\,\ldots,a_n\},\quad a_i=0,1\,.
\ee

An  important feature of  system (\ref{eq:HS_1}), (\ref{eq:HS_2})  noticed originally in \cite{Ann}
 even before this system was obtained in \cite{Vasiliev:1992av} is that it remains formally consistent
if  the fields $\WW$ and $B$ are valued in any associative algebra, for instance, in
the matrix algebra $Mat_N(C)$. As a result,
the terms corresponding to different sequences of $a_i =1$ or $0$ like $\{0,1,0,0,1,0,\ldots \}$ \etc,
 referred to as  $\va$, are separately $\dr_z$-closed.
Then the homotopy operators (\ref{pp}) are allowed to be different for different
  $\va$. The simplest option is
\be
\label{CP}
a^\va{} _{A}=c_0(\va) Y_A +\sum_{j }c_j (\va) \p_{j A}\q
  \va=\{a_1,\ldots,a_n\}  \,,
\ee
where $\p_{iA}$ is the derivative with respect to the  argument of the $i^{th}$
factor   $\Phi^{a_i}(Y_i;K)$. It is important  that the modification via
derivative homotopy shift  affects locality
when two or more  arguments are available, \ie only at the nonlinear level.

This construction provides a broad extension of the class of allowed  homotopy
operators. In fact it can be further extended by letting the coefficients $c_j(\va)$ be
 arbitrary functions of  the covariantly contracted combinations  $\p_{jA} $ and  $Y_{A}$.
Practical analysis shows however that the simplest extension with constant $c_j(\va)$
is sufficiently general. As shown in more detail in \cite{DGKV},  the class of shifted homotopy
operators with constant shift coefficients is distinguished by the property of being
closed under the elementary operations underlying the perturbative analysis like star products etc.
Let us stress that although a number of free parameters in the shifted homotopy operators
increases linearly with the order of the vertex in question, this freedom is
uncomparably  smaller
than the functional freedom in  general homotopy operators.

\section{$Z$-dominance Lemma}
\label{Locality lemma}

The following evident formula \be\label{genCCC}
 \underbrace{C(Y)*\ldots*C(Y)}_n=\exp  i \Big( \sum_j p^j{}_\ga y^\ga
 -  \sum_{j<k} p^j{}_\ga p^k{}^\ga
+\sum_j \bar p^j{}_\pa \by^\pa  -  \sum_{j<k} \bar p^j{}_\pga \bar p^k{}^\pga \Big)
C(Y_1)\ldots C(Y_n)\big|_{Y_j=0} \ee
suggests  that, to control locality,  it suffices to consider the exponential parts
 of the operators acting
 on ordered products $  C(Y_1;K )C(Y_2;K )\ldots C(Y_n;K )$
     focusing on  the  derivatives  $  p_\ga^j{} $,
 $  \bar p_\dga^j{}$ (\ref{TailorC}). To simplify  analysis it is convenient to
 define $  p^j{} $ and $  \bar p^j{}$ as respecting the chain rule
 \be\label{insensitiveK}
 p^i   ( C(Y_1;K )C(Y_2;K ))= p^i   ( C(Y_1;K ))C(Y_2;K )+  C(Y_1;K )p^i   C(Y_2;K )
\ee
in a way insensitive of the dependence of $C(Y_1;K )$ on $K$. (Formally this can be achieved
following \cite{Prokushkin:1998bq,Didenko:2017qik} by
introducing additional Clifford elements that anticommute with the Klein operators.)
For each factor of  $ C(Y_j;K )$ $  p^j_\ga{} $ is defined as the left derivative, \ie
$
  p_\ga{} (C^{ij}(Y)k^i \bar k^j )=- i\f{\p}{\p y^\ga} (C^{ij}(Y))k^i \bar k^j \,.
$

 Also it is useful to
keep track of  the extra degree of Klein operators originating from the operators $\gamma$ \eq{gamma=}.
Hence, general exponential representation for, say, the order-$n$ corrections in
the zero-forms $C$ is
\be
\sum_{\pp\bar \pp}\int d\tau P^{\pp\bar \pp}_{n}E^{\pp\bar \pp}_{n}(\tau)C(Y_1)\ldots C(Y_n)\big|_{Y_j=0}\,,
\ee
where $ P^{\pp\bar \pp}_{n}$ is some polynomial of $z,y$ and $p^i$ and their conjugates
 with coefficients being
regular functions of the homotopy parameters $\tau$, and
\be
E^{\pp\bar \pp}_{n }= E^{\pp}_{n}\bar E^{\bar \pp}_{ n}\q
\label{En} E^\pp_n(T ,A,B,P,p|z,y)=
\exp  i ( T z_\gga y^\gga - A_j p^j_\gga z^\gga- B_j p^j_\gga y^\gga
+ \half  P_{ij}  p^i{}^\gga p^j{}_\gga ) 
k^\pp\,,
\ee
where $\pp=0,1$ and parameters $T \in   \mathbb{C}$,
 $A,B  \in   \mathbb{C}^n$, $P_{ij}=-P_{ji}\in\mathbb{C}^n\times\mathbb{C}^n$
  may be $\tau$-dependent.

For instance, in \cite{Vasiliev:2017cae} it was shown that the second-order correction to
$B(Z;Y;K) $ that eventually leads to local HS equations in the zero-form sector is
 $\Hh_{cur}^{loc}=B^{loc}_{2\eta}+B^{loc}_{2\bar\eta}$ with
 \be\label{bloc0}
 B^{loc}_{2\eta}=
 \f{1}{2}\eta \int d_+^3\tau \Big (
 \delta'(1-\sum^3_{i=1}\tau_i)+iy_\ga z^\ga
 \delta (1-\sum^3_{i=1}\tau_i) \Big )\exp (X^{loc})
 C(Y_1;K)\bar*C(Y_2;K) \Big \vert_{Y_{1,2}=0}  \,k \,,
 \ee
 where $\bar*$ is the star product with respect to barred variables,
 \be
d_+^3\tau := d\tau_1 d\tau_2 d\tau_3 \theta(\tau_1)\theta(\tau_2)\theta(\tau_3) \q
\theta(\tau) = 1(0) \quad \mbox{if} \quad \tau\geq 0 (\tau<0)\,
\ee and
 \be\label{X}
X^{loc}=  i \tau_3  z_\ga y^\ga +\tau_3 z^\ga (\p_{1\ga} +\p_{2\ga}) +
y^\ga (\tau_2 \p_{2\ga}-\tau_1 \p_{1\ga}) +i\tau_3 \p_{1\ga}\p_2^\ga
 \,.
\ee
$B^{loc}_{2\bar\eta}$ is complex conjugated to $B^{loc}_{2 \eta}$.

 What we would like to explain now is that from    (\ref{bloc0}) it immediately
follows that the equations of motion in the sector of physical fields are local
rather than being {\it a priori} minimally non-local according to the argument of
\cite{Vasiliev:2017cae}.

Indeed, the first nontrivial correction to the field equations in the zero-form sector is
\be
\label{defc}  \dr C+ \go*C -C*\go+\Hh_{cur}(w,\PPP)=0\,,
\ee
where
\be
\label{JCC}
\PPP(Y_1,Y_2;K|x) := C(Y_1;K|x)  C(Y_2;K|x)\,.
\ee
Here $C$ and $\go$ are $Z^A$-independent and hence the correction $\Hh_{cur}(w,\PPP)$ must
be $Z$-independent as well. This  happens because consistency of the equations
guarantees that the corrections belong to the $\dr_Z$--cohomology. Though in practical computations
it is sometimes convenient to set $Z=0$ to simplify the derivation of
 the explicit form of the corrections to field
equations, this is not necessary since the $Z$-dependence should drop out anyway as a
consequence of the previously solved equations implying that $\dr_Z$ of the both sides of equations
is zero.

Practically, this works as follows. The $Z$-dependent term in the exponential contains
the integration homotopy parameter $\tau_3$. The fact, that the \rhs of (\ref{defc})
must be $Z$-independent implies that the integral over $\tau_3$ on the \rhs of (\ref{defc})
must reduce to integration over  such a total derivative that the final result is located
at the lower integration limit $\tau_3=0$. This however means that not only the $z$-dependent
term $i\tau_3z_\ga y^\ga$ in  the exponential in  (\ref{defc}) disappears but
the term $\tau_3 \p_{1\ga}\p_2^\ga$ must disappear as well,  so that the final expression will contain at most a finite number of contractions in the preexponential
with respect to the undotted variables, leading to a local result.

Generally, we arrive at the following {\it $Z$-dominance  Lemma}
\\\label{Lemma 1}
{\it Lemma 1:} All terms in the exponential representation
(\ref{En}) dominated by the coefficients in front of the $Z$-dependent terms $T(\tau)$ and $A_i(\tau)$
do not contribute to the field equations on the $\dr_Z$-cohomology-valued dynamical fields.

Note that hatted coefficients $\hat B_j(\tau)$ and $ \hat P_{ij} (\tau)$ in (\ref{Enh})
coincide with the $\dr_Z$ cohomology reduction of $ B_j(\tau)$ and $ P_{ij} (\tau)$
while analogous reduction of the  coefficients $T(\tau)$ and $A_j(\tau)$ in front of the
$z$-dependent terms in (\ref{En}) is zero.

This simple lemma allows us to show  that the level of non-locality of
HS equations can be decreased in higher orders  by an appropriate choice
of  shifted homotopy operators (\ref{pp}). Let us stress that $Z$-dominance Lemma 1
applies to each term in expressions containing  linear combinations  of a finite number of
exponentials  (\ref{En}) as is most easily seen by rewriting the
$z$-dependent part of the exponentials in the form
\be
\exp  i ( T(\tau) z_\gga y^\gga - A_j(\tau) p^j_\gga z^\gga) =
\int d t d t_j \delta (T(\tau)-t)\prod_j\delta (A_j(\tau) - t_j)
\exp  i ( t z_\gga y^\gga - t_j p^j_\gga z^\gga)
\ee
allowing to rewrite a sum of integrals of different exponentials as a sum of terms in the
integration
measure in front of a single exponential factor $\exp  i ( t z_\gga y^\gga - t_j p^j_\gga z^\gga)$.

\section{Pfaffian Locality Theorem}
\label{Pfaffian Locality Theorem}

Here we  prove Pfaffian Locality Theorem (PLT) stating that, in the holomorphic sector,  there exists such a choice
of the shifted  resolutions that the  matrix $P_{ij}$ of (\ref{En})
is degenerate.
In the second order this  implies that $\PP\left(\f{\p}{\p y_i}\right)=0$ and, hence,
 $J_2$ is local in agreement with
 \cite{Vasiliev:2016xui,Vasiliev:2017cae}. In higher orders  PLT implies
 at least the decrease of the level of non-locality  indicating however that
it can be decreased  further.

PLT heavily relies  on the properties of  star product \eq{star2}
and shifted homotopies. In our analysis we focus on the dependence on the zero-forms $C$
discarding the dependence on $\go$ and $\goo$ that does not affect spin locality in the $4d$ HS theory.

It is useful to  restrict the representatives of the exponential classes (\ref{En}) as follows.

    {\it Even  } class
   \bee\label{bsfield} \SSS_n:\quad
  E^\pp_n(T ,A,B,P,p|z,y)\q \pp=n|_{2}\q n\ge1
 \eee
 with parameters satisfying
\bee\label{propertS}
 \sum_{j=1}^n (-1)^j A_j=- T
\q\sum_{j=1}^n (-1)^j B_j=0
\q
\sum_{i=1}^n (-1)^i P_{ij}= B_j.
  \eee

  {\it  Odd}    class \bee
\label{bbfield}    \BS_n :\quad E^{{\mathbf{p} }}_n(T ,A,B,P,p|z,y),\qquad  {\mathbf{p} }=(n+1)|_{2}\q  n\ge 0
 \eee
 with
 \bee\label{propertB}
 \sum_{j=1}^n (-1)^j A_j=0
\q \sum_{j=1}^n (-1)^j B_j=1-T
\q
\sum_{i=1}^n (-1)^i P_{ij}=-A_j.
  \eee

 Particular cases include
  \be\label{initb}    1\in  \SSS_0 \q \kappa k\in \BS_0
 \q   \exp i(   p^1_\gga y^\gga )\in \BS_1\,, \ee
where $ \exp i(   p^1_\gga y^\gga )$
  is generated by the dynamical field $C(y,\by)$ via \eq{genCCC} with $n=1$
  while $\kappa   k$ is a part of $\gamma$ \eq{gamma=}.

Perturbative analysis   implies that corrections to dynamical equations at any perturbation order
 can be constructed inductively, starting from
$\gamma $ \eq{gamma=}, $  C(y,\by|x)$ and $ \go(y,\by|x)$
via  application of the star product,
  shifted resolutions, cohomology projectors, operators $\D_\goo $ \eq{Dkrugloe} (not affecting
locality)  and $\dr_x$. By {\it Structure Lemma 6} proven in
 Sections \ref{mapstar}-\ref{d},
with the proper choice of homotopies (\ref{vbb}), (\ref{vbs}), these operations
respect the classes $\E_n^j$.

 This
  allows  us to decrease the level of non-locality
  in the  higher   order corrections to dynamical equations in the zero-form (anti)~holomorphic sector.  Indeed, in this case
   exponential parts of the order-$n$ deformations $J^1_n(Y;K|x)$ to
  the field equations in the zero-form sector \eq{dynschnW}   are of the form (\ref{En})
with  the  parameters obeying   \eq{propertB}.
According to Lemma 1  the coefficients $T$ and $A_i$
trivialize  in the $\dr_Z$-cohomology. Hence   \eq{propertB} yields
 \be \label{propertBP}
 \sum_{i=1}^n (-1)^i P_{ij}(\tau)=0
  \ee
in the $\dr_Z$-cohomology proving  \\ {\it Pfaffian Locality Theorem}:
the shifted homotopy can be chosen  in such a way that the matrix $P_{ij}$ be degenerate
with the null-vector \eq{propertBP}
     in the (anti)holomorphic sector of the dynamical field equations in the zero-form sector.

 In even interaction orders condition \eq{propertBP} is essential.
  For instance, from Section \ref{hommap} it  follows, that to obtain a local
    form of dynamical equations   to the second  order in the holomorphic sector,
   it is necessary to take   the shifted resolution operator \be
   \hmt_{\gb y+ i\ga\p_{y_1}-i(1-\ga ) \p_{y_2}}\ee
  with arbitrary parameters $\ga$ and $\gb$.
 Details of the derivation of the local form   of equations
  are presented in  \cite{DGKV} where
  it is  also proven  that the resulting equations coincide with those of \cite{Gelfond:2017wrh}
  up to $\gb$-dependent  local field redefinitions.

In odd  orders the antisymmetric matrix $P_{ij}(\tau)$ is automatically  degenerate.
However, the additional information following from (\ref{propertBP}) is that it admits a
null vector independent of the homotopy parameters $\tau$ on which $P_{ij}(\tau)$ depends.
 Though we do not know yet whether condition \eq{propertBP} increases the degree of  degeneracy
 of $P_{ij}(\tau)$ further or not, its special structure implying that the vertex
 depends on some linear combinations of helicities associated with different fields
 suggests that the level of non-locality of the resulting vertices
 which are rather unusual from the QFT perspective is likely to be further reducible.
 We interpret this as a possible indication that   HS interactions
 may admit a spin-local form in all orders.

\subsection{Star-product  mapping}
\label{mapstar}Straightforward computation proves \\
{\it Lemma 2:}
\bee \label{prodg-eveng-odd}
\E^j_n*\E^i_m  &\subseteq&  \E^{(j+i)|_2}_{m+n}\,.\qquad\eee
The proof is by virtue of
 Eqs.~\eq{star2}, \eq{kk} which give for any    $\pp,\pp'=0,1$
   \be \nn
    E^{ \pp}_n(T ,A,B,P,p|z,y) * E^{\pp'}_{n'}(T' ,A',B',P',p'|z,y)
    = E^{ (\pp+\pp')|_2}_{n'+n}(T'' ,A'',B'',P'',p''|z,y)\,,
\ee  
with
  \bee\label{starsimpl}
 T''   &=& T (1-T' )+T' (1-T )\q
   \\ \nn
    A''{}_{j }&=&    (1-T' )  A{}_j   -T'    B{}_j \q 
       A''{}_{l'}=(-)^{ \pp} \left((1-T ) A'{}_{l'}+  T   B'{}_{l'}\right)\q 
   \\ \nn
 B''{}_j &=&   -T'   A{}_j+(1-T' )  B{}_j\q
 B''{}_{{l'}}=(-)^{ \pp} \left(  T   A'{}_{l'}+  (1-T )  B'{}_{l'}\right)
  \q
\\ \nn
 P''{}_{IJ} &=& P{}_{ij}+ P'{}_{m'l'  }-  (-)^{ \pp}\left(A{}_i+B{}_i\right)
 \left(A'{}_{l'}-B'{}_{l'}\right)
  +    (-)^{ \pp}\left(A'{}_{m'}-B'{}_{m'}\right)
 \left(A{}_j+B{}_j\right)
\eee
($i,j=1,\ldots,n;$ $m',l'=1,\ldots,n'$)
 from where Lemma 2 follows straightforwardly $\Box$

\subsection{Homotopy  mapping}
\label{hommap}
Let
 \be\label{vmupar} \mathrm{s}_n(\mu,v )=  v_j p {}^{j} 
   +\mu y\q \mu\in\mathbb{C}\,,  v\in \mathbb{C}^n
  .\ee
       The mapping 
       \be \label{hmtEn}   \hmtop_{ \gt,\shi_n(\mu,v)}\left(
 E_n(T ,A,B,P )\right)=  E_{n}(T' ,A',B',P' )\,
  \ee
     with
 \bee \label{hmtEn=}&&
 T' =\gt  T\q A'_i =\gt  A_i\q B'_i  =B_i +  (1-\gt )  T v_i
 -(1-\gt )\mu A_i
 \q\\ \nn&& P'{}_{ij}=P_{ij}+ (1-\gt )\left( A _{j} v_i-A_i v_j\right) 
  \,\qquad\qquad \eee
        results from the application of the   resolution $\hmt_a$
       \eq{homint} with $a=\shi_n(\mu,v )$
      \bee\label{homintE} \hmt_{\shi_n(\mu,v )} \left\{
\phi(z,y,p ,\theta)  E_n(T ,A,B,P )\right\}\\ \nn=\int_0^1 \f{d\gt}{\gt} (z+ \shi_n(\mu,v ))^\ga
\f{\p}{\p \theta^\ga}
 \phi( \gt z- (1-\gt) \shi_n(\mu,v ),y,p,\gt\theta) E_{n}(T' ,A',B',P' )\,,
\eee
 where $\phi(z,y,p, \theta )$ is some pre-exponential factor containing a finite
 number of $p {}^{j}$.   Elementary calculation yields \\
 {\it
 Lemma 3:}
 If
  \be \label{vbb}\sum_{j=1}^n (-1)^j v^{1}_j= 1 \,,\ee then
for any $\gt$ and $\mu$ \be\label{mappbb}
 \hmtop_{ \gt,\shi_n(\mu,v^{1})}: \BS_n \to \BS_n \,. \ee

Indeed,  if parameters $T  ,A ,B ,P $ satisfy \eq{propertB} then, for any
 $\gt$, under the
assumptions of {  Lemma 3}
 parameters $T' ,A',B',P'$ of
$\hmtop_{ \gt,\shi_n(\mu,v^{1})} \left(
E_n(T ,A,B,P )\right)$ \eq{hmtEn} can be easily shown
 to    satisfy
\eq{propertB}   by virtue of \eq{hmtEn=} and \eq{vbb}. Hence
$\hmtop_{ \gt,\shi_n(\mu,v^{1})}\left(
\BS_n(T ,A,B,P )\right)\in  \BS_n$\,$\Box$

Analogously, one proves
\\{\it
 Lemma 4:} If
   \be\label{vbs}  \sum_{j=1}^n (-1)^j v^{0}_j =-\mu\,,\ee then for  any $\gt$  and $\mu$
\be\label{mappbs}
 \hmtop_{ \gt,\shi_n(\mu,v^{0})}: \SSS_n \to \SSS_n \,. \ee

Note that if $\mu=-1$ the conditions \eq{vbb} and \eq{vbs} coincide.
This may be important in practical analysis.
 \subsection{$\dr_x$ mapping}
\label{d}
 It remains to   consider  the mapping   generated by
$\dr_x$ that acts nontrivially on the dynamical
 fields $\go$ and $C$. The action on $\goo $ does not affect locality.
 By Eq.~\eq{dynschnW}
 \bee\label{mapgodx} {\dr_x}\go (Y;K|x) {\longrightarrow}  \sum_{j }  J^0_{j }(Y;K|x) \q
 {\dr_x} C(Y;K|x) {\longrightarrow} \sum_{j} J^1_{j }(Y;K|x)\,.\eee

 Note that the lower label $j$ of $ J^i_{j}(Y)$ equals to the total degree in the dynamical fields,
while the respective $k$-equipped exponentials depend on the degree in $C$.
For the future convenience we set
\be J^i_{j }(Y):=J^i_{j_\go+ \bjc }(Y)\q i=0,1\,,\ee
where $j_\go$ and $\bjc$ are the degrees of $ J^i_{j }(Y)$ in $\go$ and $C$, respectively.
  $J^{0,1}_j $ do  not  depend  on $z$.
Hence the $k$-equipped exponential $
\widetilde{E}^{ \tilde{\pp}}_{\bjc}$ \eq{En} of $J^{0,1}_{j_\go+\bjc}(Y)$ is\be
\label{subst00}\widetilde{E}^{ \tilde{\pp}}_{\bjc}(\widetilde{B},\widetilde{P} |y):=
E^{\tilde{ \pp}}_{\bjc}(0 ,0,\widetilde{B},\widetilde{P} |0,y)\,.\ee

Since  ${\dr_x}\go (Y;K|x)$ \eq{mapgodx} contributes to the sector of two-forms it
does not affect field corrections
$B^{(n)}$ \eq{B=} and dynamical equations on the zero-form $C(Y)$
\eq{dynschnW} for which we obtain schematically
 \bee
 \label{dxPhiord}
\dr_x C(Y_1)\ldots C(Y_n) = \sum_i
C(Y_1)  \ldots\widehat{C(Y_i)}   \ldots C(Y_n)\Big|_{
C(Y_i) \to \sum_{j} J^1_j(Y_i )
} \,.
\eee

Eq.~\eq{dxPhiord}   yields
\be\label{dxPhiordE} E^\pp_n(T ,A,B,P |z,y)\stackrel{\dr_x}{\longrightarrow}\sum_{\bjc}\sum_{i }
E^{\pp^{i,\bjc}}_{n+\bjc-1}(T^{i,\bjc} ,A^{i,\bjc},B^{i,\bjc},P^{i,\bjc}|z,y).
\ee
The resulting mapping $S_{i,\widetilde{E}^{\tilde{\pp}}_\bjc}(E^\pp_n )$ generated by $\dr_x$ \eq{dxPhiordE}
for any  $i$ and $\bjc$
 is \bee \label{Subst1E}
S_{i,\widetilde{E}^{\tilde{\pp}}_\bjc(\widetilde{B},\widetilde{P},\tilde{p})}\big(E^\pp_n( T{}  ,A{} ,B{} ,P{} ,p{}  ) \big)=
E^{(\pp +\tilde{\pp})|_2}_{n+\bjc-1} (T{}' ,A{}',B{}',P{}',p{}')\q\eee where
parameters are defined straightforwardly via Eq.~\eq{En}. For instance,
\bee&&
\beee{lll}  T{}'   = T  \q&&
\\\nn
     A{}'{}_{k}=   A{}_k \q & B{}'{}_{k }=  B{}_k \qquad & \mbox{for }  k<i\q
\\ \nn
       A{}'{}_{{\tilde{k}}+i-1}= - A {}_{i} \widetilde{B}{}_{{\tilde{k}}}  \q &
     B{}'{}_{{\tilde{k}}+i-1} =- B{}_{i} \widetilde{B}{}_{{\tilde{k}}} \qquad&
     \mbox{for }1\le {\tilde{k}}\le \bjc \q
\\ \nn
    A{}'{}_{k+\bjc-1}=  (-)^{\tilde{\pp}} A{}_k \q & B{}'{}_{k+\bjc-1}=  (-)^{\tilde{\pp}} B{}_k \qquad&
    \mbox{for }  k\ge i+1\q
\eeee\\  \label{Subst11=}&&  \{p'_1,\ldots,p'_{\bjc+n-1}\}=
\{p_1,\ldots, p_{i-1},\tilde{p}_1,\ldots,\tilde{p}_{\bjc}, {p}_{i+1},\ldots, {p}_{n}\}.\eee

For odd $k$-equipped exponentials $\widetilde{E}^{\tilde{\pp}}_\bjc $ this gives   the following
\\
 {\it
 Lemma 5:}
 For any $  E^{(m+1)|_2}_m $   and any   $i\in[1,n+1]$ \be\label{mappbbS}
S_{i,E^{(m+1)|_2}_m(0 ,0,B,P,p|0,y)} : \E^a_n \to \E^a_{n+m-1} \,\q a=0,1.  
 \ee
Indeed, since $E^{(m+1)|_2 }_m\in \E^1_m $
then by virtue of   \eq{propertB}
  \bee \label{propertB0}
&&   \sum_{{n}=1}^m (-1)^{n} \widetilde{B}_{n}= 1
,\qquad
\sum_{k=1}^{m} (-1)^k \widetilde{P}_{k{n}}= 0  \,.
\qquad  \eee
Hence \eq{Subst11=} yields \bee \label{propSbbAB}
\sum_{k=1}^{m+n-1}(-)^k A{}'{}_{k}= \sum_{k=1}^{n}(-)^k A{}_k
  \q
  \sum_{k=1}^{m+n-1}(-)^k B{}'{}_{k}
  = \sum_{k=1}^{n}(-)^k B{}_k
  \qquad\eee
    satisfying  conditions \eq{propertS} for $a=0$ and \eq{propertB} for $a=1$.
Analogously, one can make sure that from \eq{propertB0} it follows that the parameters $ P{}' _{k j}$ on the \rhs of
\eq{Subst1E}
 satisfy the respective conditions \eq{propertS} for $a=0$ and \eq{propertB} for $a=1$.$\Box$

Let us stress that otherwise, if  $\widetilde{E}^{\tilde{\pp}}_\bjc $  \eq{subst00}  is even, the resulting $k$-equipped exponential
\\$E^{(\pp +\tilde{\pp})|_2}_{n+\bjc-1} (T{}' ,A{}',B{}',P{}',p{}' )$ \eq{Subst1E}
  in general  has no
definite parity.

\medskip

 By induction over perturbation orders, {Lemmas 2-5}   provide  following {\it Structure Lemma}
\\
    {\it Lemma 6}:
 If the perturbative analysis in the  one-form sector  contains shifted resolutions
 $\hmtt$ satisfying   \eq{vbs}, while that in the zero-form sector
 contains shifted resolutions
  satisfying   \eq{vbb}, then   all   $B_j$
generate  odd $k$-equipped exponentials, while  all space-time zero-  and one-form components of
$\WW_j$,  not containing terms resulting from ${\dr_x}\go (Y)$ \eq{mapgodx},  generate   even $k$-equipped exponentials  in the holomorphic sector.

Antiholomorphic sector analysis is analogous up to
swap of dotted and undotted spinors.

    \section{Conclusion}
  \label{conc}
In this paper we explain how to extend the class of homotopy operators in HS theory
to make it possible to systematically analyze locality of interactions derived from
nonlinear HS equations. It is shown that a number of available homotopy operators
increases quickly with the order of nonlinearity, containing in particular a subclass
 of homotopy operators that lead directly to the known lower-order local results as
 shown explicitly in \cite{DGKV}. Also we
 prove a {\it  $Z$-dominance Lemma} giving a sufficient condition  controlling locality of field
 equations on dynamical fields and {\it Pfaffian Locality Theorem} (PLT)  showing how to
 choose generalized homotopy operators to reach that the Pfaffian  matrix of
 derivatives acting on spinor variables of different fields in multilinear corrections degenerates.
 As shown in \cite{DGKV},  the  choice suggested by  PLT in the case of  bilinear corrections leads to the local results of
 \cite{Vasiliev:2016xui,Vasiliev:2017cae}. In the higher orders the PLT
 allows us to choose homotopy operators in such a way that
 the level of higher-order non-locality gets decreased. Indeed, PLT implies that,
  for the proper homotopy choice, {\it a priori} infinite expansion in
  spinor variables turns out to be finite with respect to at least one their linear
  combination associated with the null vector (\ref{propertBP}) of the Pfaffian derivative matrix. This result is somewhat analogous to the
 conclusions of \cite{Ponomarev:2017qab}.

 To appreciate it, the following comments have to be taken into account. The structure
 of the remaining non-local  higher-order interactions obtained by virtue of the
 homotopy operators satisfying  PLT is very special, containing
 some linear combinations of helicities associated with different fields in a vertex
 as prescribed by \eq{propertBP}.
   Such vertices are rather unusual from the QFT perspective
 and are anticipated to be further removable by an appropriate homotopy choice.
 A  related comment is that conditions of  PLT leave a lot of freedom in the choice of shifted homotopy operators in higher
 orders to be used to further reduce the level of non-locality of HS interactions.
 Hopefully,  there may exist a specific homotopy  choice leading to spin-local
 higher-order nonlinear corrections at any  order.

 The conjecture that contact HS interactions can be spin-local should not be
  interpreted as the claim that all HS interactions are space-time local.
 Most likely they are not due to the  spin-current exchange phenomenon.
 Indeed, what is proven in our formalism is that contractions with respect to spinorial
 variables of $C(Y;K|x)$ are suppressed. However, their relation to the space-time derivatives is direct only at the  level of free equations (\ref{tw}) which, however, receive nonlinear corrections at higher orders. As a result, the relation between space-time and spinor derivatives of $C(Y;K|x)$ becomes nonlinear and  also involves higher derivatives.
 Due to summation over different spins this may eventually lead to further
 $x$-space non-localities. This mechanism is somewhat analogous to the current exchange mechanism in QFT. Our results indicate that contact HS interactions may be spin-local in all orders. On the other hand at the present stage it is not  clear
 what kind of space-time non-locality is physically admissible in HS theories. The idea is first to identify the  spin-local or minimally non-local scheme in HS theory and then investigate its properties.

It should be stressed that the results of this paper are heavily based on the specific
form of HS equations (\ref{eq:HS_1}), (\ref{eq:HS_2}) and, in particular,
 of the star product (\ref{star2}). Extension of the analysis of this paper
to other cases including the sector of one-forms $\go$ and mixed (non-holomorphic) sectors needs application of the remarkable properties of the shifted homotopy
formalism elaborated in \cite{DGKV} where a number of examples of its applications are
presented.

The results of this paper, which
 are applicable not only to the $4d$ HS theory of \cite{Vasiliev:1992av} but also to $3d$ HS theory  \cite{Prokushkin:1998bq} and Coxeter HS theories proposed
  recently in \cite{Vasiliev:2018zer},
provide a  step towards complete analysis of
the level and role of non-locality in HS gauge theory. Once a spin-local or minimally
non-local formulation of the HS gauge theory is identified this will allow one
to analyze such important issues as causality and, in the framework of Coxeter
HS theory of \cite{Vasiliev:2018zer}, relation with analogous
aspects of String Theory. An extension of our results to higher orders is also
of great importance. Though some progress in that direction is reported in
\cite{DGKV} a lot more remains to be done. In particular, it would be extremely
interesting to compare predictions of the bulk HS equations against
holographic results on the simplest quartic vertices of \cite{Bekaert:2015tva,Ponomarev:2017qab,Sleight:2017pcz}.

 \section*{Acknowledgements}
We  would like  to thank Slava Didenko  and Tolya Korybut for useful discussions and also Slava Didenko for valuable comments on the manuscript.
We acknowledge  a partial support from  the Russian Basic
Research Foundation Grant No 17-02-00546.  The
research  was supported in part by the International Centre for Theoretical Sciences (ICTS) during a visit for participating in the program - AdS/CFT at 20 and Beyond.
The work of OG is partially supported by the  FGU FNC SRISA RAS (theme   ¹ 0065-2018-0004).

\end{document}